\documentclass[lettersize,journal]{IEEEtran}

\usepackage[utf8]{inputenc}
\usepackage{amsmath,amsfonts}
\usepackage{algorithmic}
\usepackage{array}
\usepackage{textcomp}
\usepackage{stfloats}
\usepackage{url}
\usepackage{verbatim}
\usepackage{graphicx}
\usepackage{xcolor}
\usepackage{glossaries}
\usepackage{steinmetz}
\usepackage{siunitx}
\usepackage{microtype}
\DeclareSIUnit{\perunit}{\text{pu}}

\ifCLASSOPTIONcompsoc
\usepackage[caption=false, font=normalsize, labelfont=sf, textfont=sf]{subfig}
\else
\usepackage[caption=false, font=footnotesize]{subfig}
\fi

\newacronym{AI}{AI}{Artificial Intelligence}
\newacronym{CPU}{CPU}{Central Processing Unit}
\newacronym{DC}{DC}{Data Center}
\newacronym{DRUPS}{DRUPS}{Diesel Rotary UPS}
\newacronym{FRT}{FRT}{Fault-Ride-Through}
\newacronym{GPU}{GPU}{Graphics Processing Unit}
\newacronym{OS}{OS}{Operating System}
\newacronym{PSU}{PSU}{Power Supply Unit}
\newacronym{ROCOF}{RoCoF}{Rate of Change of Frequency}
\newacronym{TPU}{TPU}{Tensor Processing Unit}
\newacronym{TSO}{TSO}{Transmission System Operator}
\newacronym{UPS}{UPS}{Uninterrupted Power Supply}

\newcommand{\IT}{\rm \scriptscriptstyle IT}
\newcommand{\GPU}{\rm \scriptscriptstyle GPU}
\newcommand{\CPU}{\rm \scriptscriptstyle CPU}
\newcommand{\ZIP}{\rm \scriptscriptstyle ZIP}
\newcommand{\DC}{\rm \scriptscriptstyle DC}
\newcommand{\grid}{\rm grid}
\newcommand{\UPS}{\rm \scriptscriptstyle UPS}
\newcommand{\cool}{\rm \scriptscriptstyle Cooling}
\newcommand{\burst}{\rm \scriptscriptstyle Burst}

\usepackage{balance}

\begin{document}

\title{Data Center Model for Transient Stability Analysis of Power Systems}

\author{%
  Alberto~Jim{\'e}nez-Ruiz,~\IEEEmembership{IEEE Member,} and
  Federico~Milano,~\IEEEmembership{IEEE Fellow}%
  \thanks{A.~Jim{\'e}nez-Ruiz and F.~Milano are with the School of Electrical and Electronic Engineering, University College Dublin, Belfield Campus, D04V1W8, Ireland. e-mails: alberto.jimenez-ruiz@ucd.ie, federico.milano@ucd.ie}%
  \thanks{This work is supported by Sustainable Energy Authority of Ireland (SEAI) by funding A.~Alberto~Jim{\'e}nez-Ruiz and F.~Milano under project FRESLIPS, Grant No.~RDD/00681.}%
  \vspace{-4mm}
}

\maketitle

\begin{abstract}
  The rising demand of computing power leads to the installation of a large number of \glspl{DC}. Their \gls{FRT} behavior and their unique power characteristics, especially for \glspl{DC} catered to \gls{AI} workloads, pose a threat to the stability of power systems.  To ensure its stability, it is required accurate models of the loads involved.  Here we propose a dynamic load model that properly captures the behaviour of \glspl{DC}.  Its three most defining features are the use of an \Gls{UPS} which sits between the server load and the grid, the cooling load represented by an induction motor, and a pulsing load that represents the transients caused by contemporary \glspl{DC} with significant \gls{AI} workloads.  The features of the proposed model and its impact on the dynamic performance of transmission systems are illustrated through a model of the all-island Irish transmission system and real-world data of the \glspl{DC} currently connected to this system.
\end{abstract}


\glsresetall %Reset acronyms

\section{Introduction}
\label{sec:introduction}

\subsection{Motivation}

\begin{figure}[!thb]
\centering
\includegraphics[width=0.95\columnwidth]{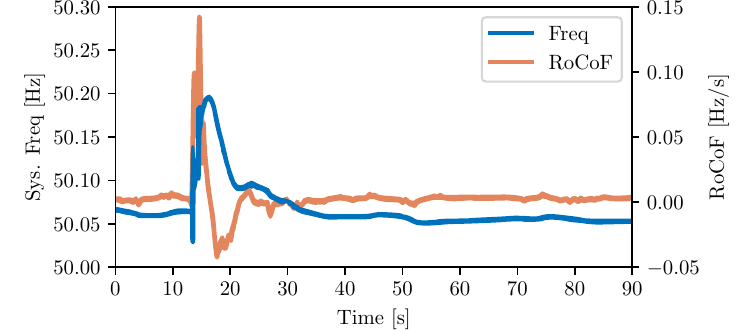}
\caption{Frequency and RoCoF for Kellystown - Woodland fault \cite{kerciImpactConverterbasedDemand2024}.}
\label{fig:Datacenter_FRT}
\vspace{-3mm}
\end{figure}

\glspl{TSO} around the world are increasingly concerned with the impact of \glspl{DC} on the dynamic performance of the system.  As a real-world example of the issues the grid with large penetration of \glspl{DC} faces, Fig.~\ref{fig:Datacenter_FRT} shows the \gls{FRT} of a \gls{DC}, where a fault caused a \qty{204}{\MW} drop of \gls{DC} demand, causing a \gls{ROCOF} of \qty{0.12}{Hz/s} and a zenith of 50.22 Hz in the all-island Irish transmission system \cite{kerciImpactConverterbasedDemand2024}.  

Their large demand also prevents \glspl{DC} from being installed in arbitrary locations, as the \gls{TSO} must assess whether they are within a constrained or unconstrained region of the electricity system, if the \gls{DC} is capable of providing flexibility in their demand by reducing consumption, and if they can bring dispatchable generation equivalent to or greater than their demand \cite{cru_2021}.

A precise dynamic model of \glspl{DC} adequate for transient stability analysis is not currently available.  \glspl{TSO} generally use generic load models which do not properly represent the dynamic behavior of \glspl{DC}.  
This work addresses this gap and proposes a dynamic \gls{DC} model that is adequate to reproduce real-world observations and anticipate potential instabilities (e.g., flapping) that the switching logic and protections of \glspl{DC} can cause to the transmission system.

\subsection{Literature Review}

There are in the literature steady-state \gls{DC} models, which are adequate for operation and economic analyses. For example, \cite{cupelliDataCenterControl2018} focuses on the thermal side of \glspl{DC} and studies the optimal response of the \glspl{DC} depending on the electricity market load and the \gls{DC} load.   Reference \cite{luoImpactDatacenterLoad2020} also studies the optimal grid placement of \glspl{DC} in the grid.  Reference  \cite{fanPowerProvisioningWarehousesized2007} discusses how to optimize the energy consumption of \glspl{DC} by using multiple workloads and optimize resource usage.  Reference \cite{zhangUnlockingFlexibilitiesData2025} suggests using the \gls{UPS} batteries to reduce the energy cost by storing energy in valley hours.

A fair amount of articles cover long-term dynamics of \glspl{DC} and model exclusively the heat transfer process and thermodynamics aspects such as finding out the best cooling strategy \cite{habibikhalajEnergyEnvironmentalEconomical2016}; the new advances in cooling technology \cite{chuResearchStatusDevelopment2023}; how to distribute the servers to enhance airflow through the \gls{DC} \cite{beghiModellingControlFree2017, anDynamicCouplingRealtime2022, fuEquationbasedObjectorientedModeling2019}; and the cooling model for long-term simulations \cite{hamSimplifiedServerModel2015}.  These studies focus on the \gls{DC} rather than on the dynamic interaction of the \glspl{DC} with the grid.

On a shorter time scale, \cite{drovtarUtilizingDemandResponse2025} proposes a dynamic load model for voltage and reactive power control suitable for \glspl{DC}, whereas \cite{suryanarayanaSystemModelingFault2021} and \cite{sunDynamicModelConverterBased2022} consider the \gls{DC} model the behavior of the server from the power electronics point of view.  The authors model the \gls{UPS} focusing on the internal dynamics of the \gls{DC} and regarding the load of the servers as dependent exclusively on their instantaneous load.  References \cite{dayarathnaDataCenterEnergy2016}, \cite{cheungSimplifiedPowerConsumption2018} and \cite{ahmedElectricalEnergyConsumption2019} approach \gls{DC} modelling from a bottom-up approach, where storage units, \glspl{CPU}, network elements, and other equipment are considered.  Again, these works focus on the device rather than on the grid.  As we show in this paper, however, \glspl{DC} are capable of fast transients and operations that cause concerns to system operators.

% The need of adapting and properly regulating the load of \glspl{DC} will thus play an important role in the foreseeable future. 

In industry, \gls{DC}'s power consumption is usually represented with a standard PQ or voltage dependent load.  These generic models do not include the protections of the \glspl{DC} that protect the servers by disconnecting the entire load from the grid.  Disconnection and \gls{FRT} logics are usually modeled \textit{ad hoc} by means of if-then conditions on voltage and frequency levels.  However these logic do not include ramps, reconnection delays and internal \gls{DC} dynamics.

Finally, in \cite{pourbeikAggregateDynamicModel2019}, the DER\_A model has been utilized to overcome some of the issues above.  This model aggregates the behavior of many individual generators (in this case utilized with negative power to emulate the load behavior) that may, after a fault, restart generating back to the grid.  After a fault, the DER\_A model reconnects part of the power consumption instantaneously, not considering any delay, which is not how \glspl{DC} usually operate.  Furthermore, this model includes terms for voltage and frequency control, which are services that \glspl{DC} do not currently provide.

\subsection{Contributions}

We propose in this article a detailed model of a \gls{DC} that takes into account its dynamic behavior, including both its unique load profile for \gls{AI} loads and their reconnection logic.  The specific features of the dynamic behavior of \glspl{DC} that we consider in this work are as follows.
\glsreset{DC}
\begin{itemize}
\item A \gls{DC} model that takes into account its cyclical load consumption for \gls{AI} \glspl{DC}.
\item A \gls{DC} model with proper disconnection and reconnection logic that represents their \gls{FRT} behavior against faults. 
\end{itemize}

The case study considers a real-world grid, namely the all-island Irish transmission system, as well as recent data and observations for EirGrid, the Irish TSO.  The Irish system is a particularly interesting case due to its size, islanded nature and the high penetration of \glspl{DC}, which is currently 800 MW over a peak of 7 GW and is expected to increase to 1.6 GW or, equivalently, 30\% of the demand by 2030. 

\subsection{Organization}

The remainder of the paper is structured as follows.  Section \ref{sec:load_classification} shows the different power consumption patterns of \glspl{DC} depending on the workload type.  Section \ref{sec:datacenterstructure} gives a general view of the internal components of  \glspl{DC} from the electrical point of view and introduces their equations, while Section \ref{sec:case_studies} shows the \gls{DC} model in action against certain case studies.  Section \ref{sec:conclusions} draws conclusions and outlines future work.

\section{Data Center Classification} 
\label{sec:load_classification}

The power usage pattern of each \gls{DC} depends heavily on the type of computation the devices of the \gls{DC} carry out.  In this article we have identified three usage patterns for the short term load demand of \glspl{DC}.

\subsection{Constant Load Pattern}

The computing power needs of the society are increasing year after year, but limitations in the characteristics of individual processors and storage units mean that increasing  \gls{DC} capacity involves scaling horizontally, i.e., increasing the number of servers. The global workload of the \gls{DC} is partitioned and assigned to each individual server so that the computing usage of each individual server is equal. As an example, \glspl{DC} owned by content delivery network companies replicate clients' data across \glspl{DC} all over the world for both large reliability and low response time.  The load demand depends on the total national or global demand of these services.  For this reason, their power consumption is regarded as constant for transient stability analysis. This pattern of consumption can be found in \glspl{DC} whose load does not vary or whose load vary very slowly throughout the day.

\subsection{Batched Load Pattern} 
\label{seq:batched_load}

\glspl{DC} owned by large research institutions, universities, or any institution which requires to compute large amounts of data in an irregular way, show a batched load pattern.  Tasks to be run on \glspl{DC} are queued and eventually executed if enough computational resources are available. 

\begin{figure}[!htb]
\centering
\includegraphics[width=\linewidth]{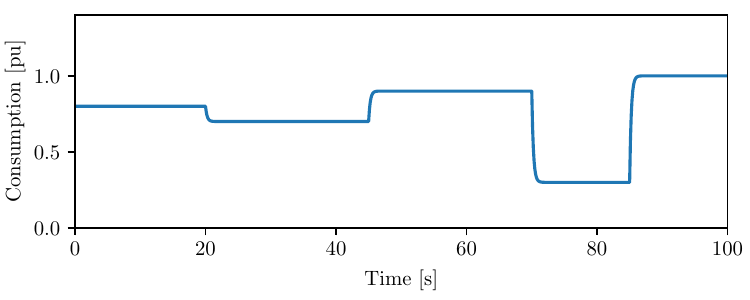}
\caption{Batched load.}
\label{fig:load_type_batched}
\vspace{-3mm}
\end{figure}

\begin{figure}[!htb]
\centering
\includegraphics[width=\linewidth]{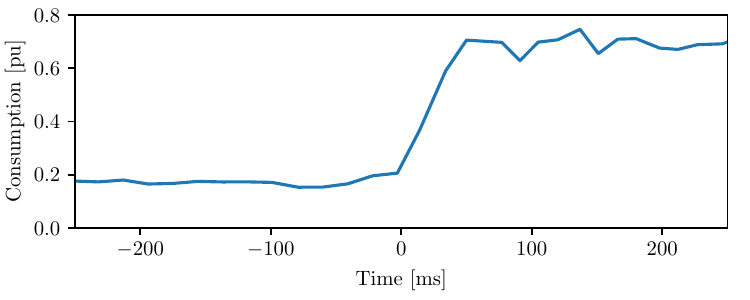}
\caption{Power consumption of transient CPU scaling driver: \texttt{intel\_pstate}.}
\label{fig:power_transients}
\vspace{-3mm}
\end{figure}

Fig.~\ref{fig:load_type_batched} plots the load of a \gls{DC} which follows this pattern. Each task runs for an irregular amount of time, and the load abruptly changes after it finishes and a new task is assigned to the \gls{DC}. The power consumption of a device that uses switching gates, such as a \gls{CPU} or a \gls{GPU} is directly proportional to its frequency and to the square of its voltage \cite{chandrakasanMinimizingPowerConsumption1995}.  These frequency changes are not instantaneous, as the \gls{OS} regulates the frequency of the \glspl{CPU}.  Fig.~\ref{fig:power_transients} shows, for example, a computer with the \texttt{Intel(R) Xeon(R) CPU E3-1245 v5 80W} \gls{CPU} that was stressed with the \texttt{stress-ng} utility in order to fully utilize all its \gls{CPU} cores.  The \gls{OS} then gradually scales up the frequency and voltage of the \gls{CPU}, thus leading to the consumption spike. 

\subsection{AI Load Pattern}

\begin{figure}[!htb]
	\centering
	\includegraphics[width=\linewidth]{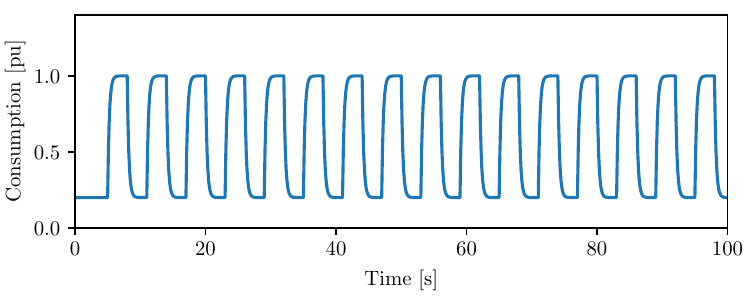}
	\caption{\gls{AI} load.}
	\label{fig:load_type_ai}
\end{figure}

The surge of \gls{AI}-based solutions has dramatically changed the power consumption patterns of \glspl{DC}.  The vast majority of \gls{AI} solutions are based on deep learning with neural networks, where their training is usually performed by feeding training examples into the neural network. This operation is repeated again and again until we achieve the optimal result.  A step that covers all the training examples is called an epoch \cite{russellArtificialIntelligenceModern2016}.  This training is performed on architectures with large parallelism potential such as \glspl{GPU} and \glspl{TPU}.  Before running the epoch training stage, data has to be transferred to the \gls{AI} accelerators, and after them the results have to be aggregated. Furthermore, epochs are usually fractioned due to the large size of the training data. This regular transition between idle \gls{AI} accelerators and usage of \glspl{GPU} and \glspl{TPU} creates a load pattern similar to the one shown in Fig.~\ref{fig:load_type_ai}.  This type of load is characterized by quick bursts of power consumption \cite{ryanAssessmentLargeLoad2025}, followed by a brief cool down period.  Training of neural networks requires that all servers work in lockstep, therefore all \glspl{GPU} and \glspl{TPU} activate and deactivate at the same time. As Section \ref{seq:batched_load} mentions, power consumption does not change instantly, and it has been likewise been smoothed out with a low pass filter.

\section{Datacenter internal structure} 
\label{sec:datacenterstructure}

\begin{figure}[!htb]
	\centering
	\includegraphics[width=0.99\columnwidth]{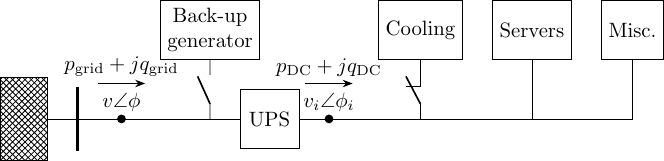}
	\caption{\gls{DC} internal structure.}
	\label{fig:Datacenter_GeneralStructure}
\end{figure}

From the electrical point of view, a \gls{DC} can be represented as shown in Fig.~\ref{fig:Datacenter_GeneralStructure}. Its components are the server load, the cooling load, the miscellaneous load, and the \gls{UPS}.

\subsection{Servers}

The majority of the power consumption of the building is done by the servers installed in the facilities (also named as IT load).  While the external grid supplies AC power, these electronic loads work with DC power. The conversion is usually performed by the \glspl{PSU} installed on each individual rack server installed within the \gls{DC}.  New types of electrical power delivery solutions which involve direct DC power delivery to the server racks are being developed that reduce the power conversion stages and increase the overall efficiency and reliability of the \gls{DC} \cite{chenDataCenterPower2023}.  Regardless of the \gls{PSU} type of these servers, they can be modeled as two distinct loads with different behavior: the \gls{CPU} and \gls{GPU}.  

According to \cite{barrosoDatacenterComputerDesigning2019}, the CPU power consumption can be modeled as:
\begin{equation}
    \label{eq:powerserverconsumption}
	p^{\rm raw}_{\CPU} = p_{\CPU}^{0\%} + \left(p_{\CPU}^{100\%}-p_{\CPU}^{0\%}\right) \, u_{\CPU} + p_{\burst} \, ,
\end{equation}
where $u_{\rm CPU}$ is the percentage of the server resources used and $p_{\CPU}^{0\%}$ and $p_{\CPU}^{100\%}$ are the power consumption when the server is idle and at full capacity respectively.  $p^{\rm raw}_{\CPU}$ represents the \gls{CPU} power consumption at $u_{\CPU}$ load, where $u_{\CPU} \in [0,1]$.  Variable $p_{\burst}$ was added to model cyclical sudden increases in power consumption in traditional \glspl{DC} \cite{kerciImpactConverterbasedDemand2024}.  This behavior can be formally expressed as:
\begin{equation}
   \label{eq:pulsetrainCPU}
   p_{\burst} = 
   \begin{cases}
      \hat{p}_{\burst}, & t \in [n\hat{T}_{\burst},n\hat{T}_{\burst}+\tau_{\burst}], n \in \mathbb{N} \, , \\
      0, &\text{otherwise} \, ,
   \end{cases}
\end{equation}
where $\tau_{\burst}$ is the duration of each burst, generally of the order of few seconds; $\hat{T}_{\burst}$ is the period of the observed transients, which are in the range of minutes, and the power increase during the spikes is $\hat{p}_{\burst}$. 

Equation \eqref{eq:powerserverconsumption} represents the power consumption of the \gls{CPU} due to the fact that the kernel or the internal \gls{CPU} performance scaling driver can gradually change the frequency and the voltage of the CPU.  The value of $u_{\CPU}$ is constant if the \gls{DC} to be simulated follows a constant load pattern or an AI load pattern.  For batched load patterns $u_{\CPU}$ shall be modeled as a piecewise function.  The power consumption and length of each interval depends on the processes being run and can be arbitrary.  In particular, in the case study, we use the jumps-diffusion process described in \cite{gudrun:2019}:
\begin{equation}
   d u_{\CPU} = c_{\CPU} \, dJ_t \, ,
\end{equation}
where $d$ indicates the differential operator, $c_{\CPU}$ is a constant and $J_t$ is a compound Poisson process, which generates a sequence of Poisson distributed jumps occurring at times $t_k$, $k \in \mathbb{N}^+$ with amplitudes that can have an any distribution.  For the simulations, we have used uniformly distributed amplitudes.
Finally, it should be noted that \eqref{eq:powerserverconsumption} regards the consumption of the \gls{CPU}, storage devices, and other auxiliary devices within the server as the aggregated \gls{CPU} load.   

Likewise, if the servers within the \gls{DC} include \glspl{GPU} or other \gls{AI} accelerators, its consumption will be equal to:
\begin{equation}
    \label{eq:powerserverconsumptionGPU}
	p^{\rm raw}_{\GPU} = p_{\GPU}^{0\%} + \left(p_{\GPU}^{100\%} - 
    p_{\GPU}^{0\%}\right) u_{\GPU} \, ,
\end{equation}
where $p_{\GPU}^{0\%}$, $p_{\GPU}^{100\%}$ and $p_{\GPU}^{\rm raw}$ mirror their \gls{CPU} counterparts. $u_{\GPU}$ shall be modeled as a train pulse:
\begin{equation}
  \label{eq:pulsetrainGPU}
  u_{\GPU} = 
  \begin{cases}
    u^{\max}_{\GPU}, & t \in [n\hat{T}_{\GPU},n\hat{T}_{\GPU}+\tau_{\GPU}], n \in \mathbb{N} \, , \\
    u^{\min}_{\GPU}, &\text{otherwise} \, .
  \end{cases}
\end{equation}
whose valleys $u^{\min}_{\GPU}$ and peaks $u^{\max}_{\GPU}$ correspond to the usage of the \gls{AI} accelerators when they are idle and active, respectively.  The duration of each pulse $\tau_{\GPU}$ is equal to the time  the \gls{AI} accelerators are running, while the period $\hat{T}_{\GPU}$ of \eqref{eq:pulsetrainGPU} is equal to the time it takes the results after each epoch to be aggregated with the \gls{AI} accelerators idle plus $\tau_{\GPU}$.

To account for transients and the fact that each server's load might be unequal and/or its response not properly synchronized, \gls{CPU} and \gls{GPU} loads of the server are smoothed out with fast low-pass filters:
\begin{align}
    \label{eq:pin_it_smooth}
	T_{\CPU}\, p'_{\CPU} &= p_{\CPU}^{\rm raw} - p_{\CPU} \, , \\
    \label{eq:qin_it_smooth}
	T_{\GPU}\, p'_{\GPU} &= p_{\GPU}^{\rm raw} - p_{\GPU} \, ,
\end{align}
where $p_{\CPU}$ represents the power consumption of \gls{CPU}-related processes (i.e., content distribution, database handling), and $p_{\GPU}$ represents processes which run on \gls{AI} accelerators, mainly \glspl{GPU} but also \glspl{TPU} and other customized devices. $T_{\CPU}$ and $T_{\GPU}$ are time constants that smooth out the sudden variations of power consumption.  As Fig.~\ref{fig:power_transients} shows, it takes $\approx 50 {\rm ms}$ for the server to reach the new power consumption state.  The transient speed can be adjusted so as to either enhance performance during bursts of \gls{CPU} usage or improve energy usage.  

Finally, the sum of CPU and GPU loads constitutes the total load of the servers, as follows: 
\begin{equation}
  \label{eq:pinserver}
  p_{\IT} = p_{\CPU} + p_{\GPU} + \eta_{\IT} \, ,
\end{equation}
where $\eta_{\IT}$ is an Ornstein-Uhlenbeck process with zero mean and bounded standard deviation \cite{rafael:2013}:
\begin{equation}
  \label{eq:noise}
  d \eta_{\IT} = -a_{\IT} \, \eta_{\IT} \, dt + b_{\IT} \, dW_t \, ,
\end{equation}
where $a_{\IT}$ is the mean reversion speed, $b_{\IT}$ is the diffusion coefficient and $W_t$ is a Wiener process.  Details on the integration of \eqref{eq:noise} are provided in \cite{rafael:2013}.

\subsection{Cooling}

The processing units of these servers dissipate heat. This thermal energy must be removed from the \glspl{CPU} and \glspl{GPU} so that they can perform their tasks at maximum performance without suffering damages. This heat reducing can be performed by the servers themselves by means of fans and chillers and, in addition, \gls{DC} buildings also offer centralized solutions such as liquid to liquid cooling, where a refrigerant is cooled down by a centralized chiller and delivered to individual server cages; and air containment strategies, which direct cooling air to the equipment. Centrifugal chiller systems perform this process by means of a compression cycle which transports the heat from the inside to the outside of the \gls{DC} \cite{oliveiraComparativePerformanceAnalysis2019}. 

The cooling load accounts for about 30\% of the total power consumption of standalone \glspl{DC} \cite{zhangSurveyDataCenter2021} and is modeled as an induction motor owing to their widespread use in compressors and water pumps  \cite{hasanuzzamanEnergySavingsEmissions2011}.  The equations that govern these motors are well known and can be found in books such as \cite{kundurPowerSystemStability1994}. The dq-axis squirrel-cage induction motor model with inclusion of stator and rotor flux dynamics was used to simulate the electro-mechanical transients that manifest after reconnection of the \gls{UPS} following a cleared fault.  The mechanical load of the motor was considered to be constant for studying short-term power system oscillations.

Note that, as the focus on this work is on short-term dynamics and the switching logic of \glspl{DC} following a system event, we do not consider temperature dynamics nor the thermodynamic model of the \gls{DC}.

\subsection{Miscellaneous}

There are elements apart from the servers and the cooling load that also consume power.  These miscellaneous loads are, for instance, lighting, video surveillance and any auxiliary element used by the staff that attends the \gls{DC} building.  They should only consider the consumption of devices associated, or installed within the facilities of the \glspl{DC}. 

The rest of the loads of the \gls{DC} are modeled as an aggregated ZIP load, as follows:
\begin{equation}\label{eq:poutZIP}
	\begin{split}
		p_{\ZIP} &= p^o_{\ZIP} \, (a_{p} + b_{p} \, v_i + c_{p} \, v_i^2 ) \, , \\
		q_{\ZIP} &= q^o_{\ZIP} \, (a_{q} + b_{q} \, v_i + c_{q} \, v_i^2 ) \, ,
	\end{split}
\end{equation}
where $p^o_{\ZIP}$ and $q^o_{\ZIP}$ are the ZIP load consumption at rated voltage; the coefficients satisfy the conditions $a_{p} + b_{p} + c_{p} = 1$ and $a_{q} + b_{q} + c_{q} = 1$; and $v_i$ is in the internal voltage magnitude of the \gls{DC}, i.e., the voltage that the \gls{UPS} sets or the voltage of the grid depending on its topology (see Section \ref{subsec:ups}).

In this work, we focus exclusively on \glspl{DC} as standalone facilities.  Mixed-use buildings, such as an office building with a floor dedicated to a server farm might consider for the rest of the building an additional dedicated model that properly features its static and dynamic behavior.

\subsection{Back-up Generator}

The back-up generator is utilized in emergency conditions, that is, only if the power outage extends through time.  During a fault, the cooling load, which is usually not protected by the \gls{UPS} and connected directly to the grid, is disconnected from the grid as well, effectively removing the entire \gls{DC} power consumption. If the outage persists, the back-up generator is started up and connected to support the \gls{UPS} and the cooling motors.  As it can be regarded as a power source that seamlessly takes over the \gls{UPS} battery, the back-up generator is not explicitly modeled here and its contribution is implicitly embedded in the \gls{UPS}.

\subsection{UPS} 
\label{subsec:ups}

\glspl{DC} are equipped with an \gls{UPS} that sits between the IT load loads and the external grid (see Fig.~\ref{fig:Datacenter_GeneralStructure}).  This \gls{UPS} can be centralized, meaning that a single device protects the entire facility, while decentralized \gls{UPS} consist of many smaller \glspl{UPS} that are installed at rack level. 

\begin{figure}[!htb]
	\centering
	\subfloat[][UPS offline mode.]{\label{fig:ups_type_a}
		\centering
		\includegraphics[scale=0.8]{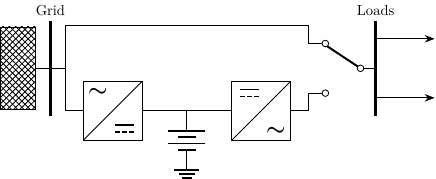}}
	
    \subfloat[][UPS online mode.]{\label{fig:ups_type_c}
		\centering
		\includegraphics[scale=0.8]{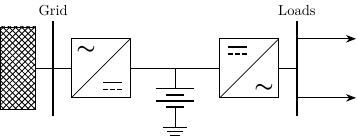}}
	
    \subfloat[][\gls{DRUPS}.]{\label{fig:ups_type_d}
		\centering
		\includegraphics[scale=0.8]{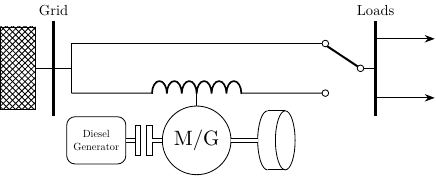}}
	\caption{UPS topologies.}
	\label{fig:ups_topologies}
\end{figure}

The main function of the \gls{UPS} is to provide the servers with power in case of a fault and thus ride through voltage disturbances on the electric grid.  Fig.~\ref{fig:ups_topologies} shows the main UPS topologies for \glspl{DC}, where Fig.~\ref{fig:ups_type_a} represents the offline mode of the UPS, where the protected loads are connected under normal operation, and they are quickly disconnected by the static switch if a fault is disconnected, to be then supplied directly by the \gls{UPS} inverter. 

Fig.~\ref{fig:ups_type_c} shows the online mode, where the loads are never connected directly to the grid, and are fed exclusively by the \gls{UPS} inverter. This topology offers the highest security out of all \gls{UPS} topologies at the expense of the additional power losses of the inverter and rectifier of the \gls{UPS}. 

The final topology shown corresponds to the \gls{DRUPS} scheme (see Fig.~\ref{fig:ups_type_d}).  A flywheel stores kinetic energy which can be transformed by a generator immediately should there a power outage be.  Under normal operation, this generator works as a motor and accelerates the flywheel.  If more energy is needed, a backup diesel generator can generate the energy required to power the loads of the system.  The generator is attached to the motor through a clutch as it is inactive under normal operation. 

From the modeling point of view, the active and reactive load that it is being protected by the \gls{UPS} and back-up generator can be written as:
\begin{equation}
  \label{eq:pqoutUPS}
  \begin{split}
    p_{\DC} &= p_{\cool} + p_{\ZIP} + p_{\IT} \, , \\
    q_{\DC} &= q_{\cool} + q_{\ZIP}\, , 
  \end{split}
\end{equation}
where the $p_{\cool}$ and $q_{\cool}$ are the active and reactive power drawn by the cooling motor.

The \gls{UPS} performs 3 main actions:
\begin{itemize}
	\item Deliver power to the loads downstream.
	\item Protect the loads during a fault by isolating them from the grid.
	\item Charge the energy storage elements of the \gls{UPS}.
\end{itemize}

We identify 3 different modes of operation that characterize the power drawn by the \gls{UPS} from the grid, namely $p_{\grid}$ and $q_{\grid}$, as follows.
\begin{itemize}
\item \textbf{Normal mode:} The loads are connected to the \gls{UPS}, but internally, the loads of \glspl{UPS} in offline mode are connected directly to the grid, while for \glspl{UPS} working in online mode the \gls{UPS} itself draws the power to feed the loads from the grid:
\begin{equation}\label{eq:pinUPS_1}
    \begin{split}
	p_{\grid} &= (1+\beta) \, p_{\DC} \, , \\
	q_{\grid} &= q_{\DC} \, ,
    \end{split}
\end{equation}
where the coefficient $\beta$ represents the losses of the \gls{UPS}.  During normal operation, electrical power is transferred through the \gls{UPS} that work with an online topology. For offline or \gls{DRUPS} modes, $\beta=0$.
\item \textbf{Emergency mode:} The \gls{UPS} supplies the loads, thus effectively disconnecting the \gls{DC} from the grid.  In this operating condition, the following constraints hold:
\begin{equation}
    \label{eq:pinUPS_2}
    \begin{split}
	p_{\grid} &= 0 \, , \\
	q_{\grid} &= 0 \, .
    \end{split}
\end{equation}
\item \textbf{Charging mode:} After a fault the \gls{UPS} replenishes the energy supplied to the loads:
\begin{equation}
   \label{eq:pinUPS_3}
   \begin{split}
      p_{\grid} &= (1+\beta) \, p_{\DC} + p_{\UPS} \, , \\
      q_{\grid} &= q_{\DC} \, .
   \end{split}
\end{equation}
where $p_{\UPS}$ is the additional power drawn from the grid to charge the battery of the \gls{UPS}.  This parameter is defined based on charging time and/or battery life considerations.
\end{itemize}

The aggregated energy stored in the \gls{UPS} and the back-up generator ($e$) varies depending on the difference between output and input power:
\begin{equation}
\label{eq:pbattery}
e' = p_{\grid} - (1+\beta) \, p_{\DC} \, .
\end{equation}
During normal mode, the right hand side of \eqref{eq:pbattery} is equal to $0$, under emergency mode equal to $-(1+\beta) \, p_{\DC}$, and to $p_{\UPS}$ under charging mode. In charging mode, when the value of $e$ reaches $e_{\rm max}$, the \gls{UPS} switches to normal mode.

\subsection{UPS Disconnection and Reconnection Logic}

The \gls{UPS} measures two quantities to switch to emergency mode: frequency and voltage. A threshold for each magnitude is defined where $f_{\rm min}$, $f_{\rm max}$, $v_{\rm min}$ and $v_{\rm max}$ are the minimum and maximum values of the frequency and the voltage, respectively.  The margins of frequency and voltage usually applied are $[0.2,0.3]$ Hz and $10\%$ of the nominal voltage.
If either of these two values is out of range the \gls{UPS} should disconnect the loads. The expression that shows this logic is:
\begin{equation}
\label{eq:disconnect}
\begin{aligned}
\mathrm{Disconnect~UPS} = & (\Delta f < f_{\rm min}) \lor (\Delta f > f_{\rm max}) \lor \\
& (\Delta v < v_{\rm min}) \lor (\Delta v > v_{\rm max}) \, .
\end{aligned}
\end{equation}
This disconnection should occur as fast as possible with no delays, unlike the reconnection.

After the fault is cleared and the bus voltage magnitude returns to an acceptable value, the \gls{UPS} can reconnect the \gls{DC} to the grid, according to any of the following schemes.
\begin{itemize}
\item \textbf{Instant reconnection:} the load is reconnected immediately after the \gls{UPS} senses the grid has returned to normal conditions. If the magnitudes are not stabilized, there is a risk of flapping, where the sudden reconnection of the load causes a drop in frequency and/or voltage below the limits of the \gls{UPS}. For this reason, a delay is usually introduced for reconnecting the loads.
\item \textbf{Delayed reconnection:} to fix flapping a reconnection delay is introduced. The typical delay ranges from $10$ to $50$ seconds. During that delay no issue with the grid must be observed.
\item \textbf{Disturbance counting:} additionally, a disturbance counting scheme might be applied. This scheme deactivates the reconnection if a certain number of disturbances are seen within a certain time. The typical number is three disturbances and the typical duration is a minute \cite{nercincident2025}.
\item \textbf{Manual reconnection:} certain \gls{UPS} systems such as the \gls{DRUPS} topology will not transfer the load back to the grid automatically, but they have to be manually reconnected by an operator \cite{nercincident2025}.
\end{itemize}

\begin{figure}[!htb]
	\centering
	\includegraphics[width=0.95\columnwidth]{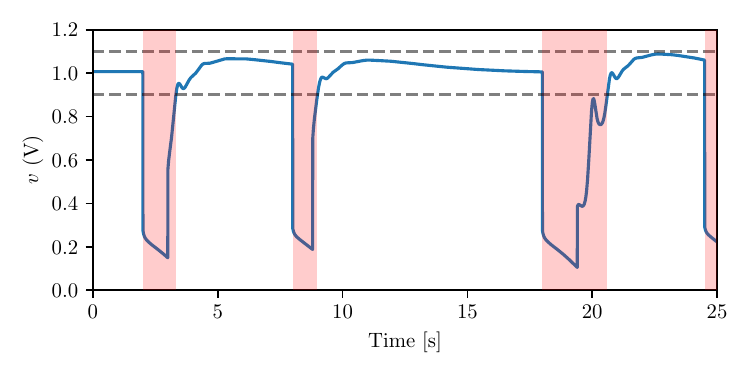}
	\caption{UPS reconnection.}
	\label{fig:UPS_reconnection}
    \vspace{-3mm}
\end{figure}

The \gls{DC} model proposed in this article counts a disturbance when the state of the UPS changes from normal to emergency mode. The time from last fault, however takes as a reference the elapsed time since the voltage and frequency parameters of the grid are under the safety margins. Fig.~\ref{fig:UPS_reconnection} shows an example of a fault that causes a voltage dip. The shaded areas represent the time intervals that, according to the \gls{UPS}, warrant a disconnection. The actual reconnection will not occur until the conditions of each reconnection scheme presented in the paragraphs above are met.

\subsection{UPS internal voltage and angle}

In normal operation, the \gls{UPS} is connected directly to the grid, thus:
\begin{equation}
v_i \angle{\phi_i} = v\angle{\phi} \, ,
\end{equation}
where $v\angle \phi$ is the external voltage at the point of connection to the grid and $v_i \angle{\phi_i}$ the internal voltage of the \gls{DC} (see Fig.~\ref{fig:Datacenter_GeneralStructure}).

After a fault, the \gls{UPS} and the back-up generators of the \gls{DC} supply power to the loads in isolation.  It is thus their sole responsibility to maintain the internal voltage of the \gls{DC} electrical installation.

During a fault, the online \gls{UPS} maintains $v_i$ without variation with respect to the normal operation.  However, offline \gls{UPS} and \gls{DRUPS} quickly isolate the loads and set $v_i$.  There are two main schemes, as follows.
\begin{itemize}
\item $v_i = 1.0$ pu.  This scheme provides the nominal voltage to the loads during the fault.  However, after reconnection, the grid voltage might have, or might have had, a different value.  This sudden voltage change might cause transients.
\item $v_i = v(t_{\rm fault}-\delta)$. Under this scheme the \gls{UPS} supplies the measured voltage of the grid after the fault.  The voltage right before the \gls{UPS} disconnect may be outside the safety margins configured by the \gls{UPS}.  Therefore, the voltage used  of $v$ is the one measured a small time $\delta$ before detecting the fault. 
\end{itemize}

The internal voltage angle $\phi_i$ is immaterial when the \gls{DC} is disconnected from the grid, as no loads of the \gls{DC} depend explicitly on the voltage phase angle.  However, to reconnect the \gls{DC} to the grid the \gls{UPS} must adjusts $\phi_i$ to match the phase $\phi$ of the external grid.  Thus, we assume $\phi_i=\phi(t_{\rm reconnection} - \delta)$ for the reconnection phase angle condition.

\section{Case Study} 
\label{sec:case_studies}

\begin{figure}[!htb]
\centering
\includegraphics[width=0.4925\columnwidth]{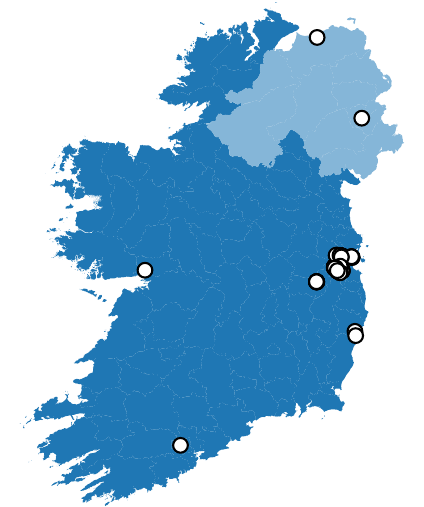}
\includegraphics[width=0.4925\columnwidth]{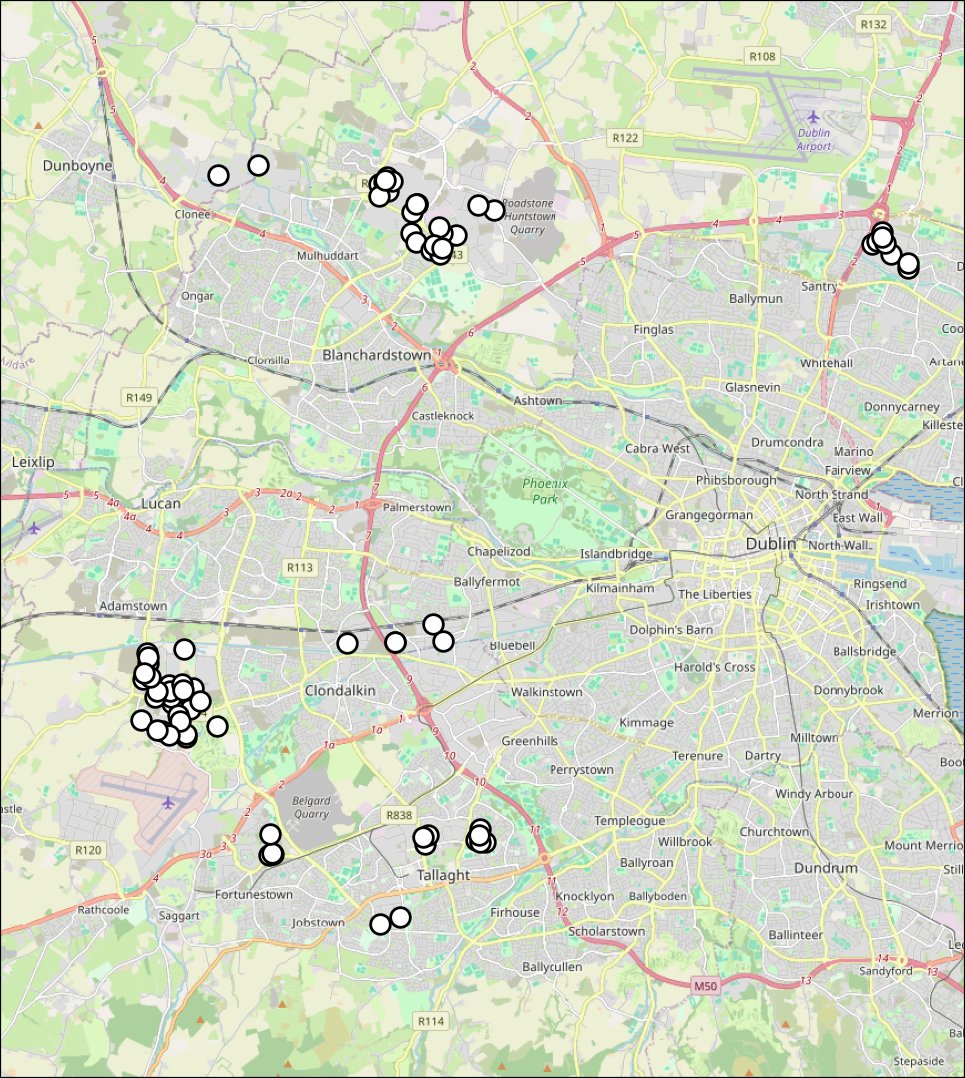}
\caption{Datacenter locations in Ireland and Northern Ireland (left) and in Dublin county (right) \cite{DataCenterMap}.  Source of road map data: OpenStreetMap.}
\label{fig:Datacenter_Map}
\end{figure}

The \gls{DC} loads are for the most part concentrated in industrial complexes of large population centers.  Fig.~\ref{fig:Datacenter_Map} shows the location of the \glspl{DC} located in Ireland and Northern Ireland (left panel) and in the Dublin county (right panel) \cite{DataCenterMap}.  Out of all the 104 \glspl{DC} located on the Irish island, 86 of them are located in the Dublin metropolitan area and, as it can be inferred from these figures, \glspl{DC}' are not only located in industrial complexes, but within these industrial areas their facilities are adjacent to each other. Thus, their power usage strains a select few transmission lines and busbars. Modeling their behavior is useful for \glspl{TSO} in order to assess the stability of the grid. 

\glspl{DC} have grown in size dramatically for the last years.  Due to economy of scale, it is incentivized to concentrate computational resources in a single facility, giving rise to hyperscale \glspl{DC}, which are buildings designed to process huge amounts of data.  Today’s largest hyperscaler sites are multiple buildings with total power of \qty{200} MW to \qty{1} GW \cite{linAdaptingDatacenterCapacity2023}.  Therefore, the \glspl{DC} considered for the case studies are large and single loads of hundreds of MWs, and not 
distributed ones.

As starting point, we have utilized a dynamic model of the all-island Irish power system that includes 1479 buses, 1851 transmission lines and transformers, 22 synchronous generators, along with their appropriate control systems, 169 wind power plants and 245 loads.  All wind power plants are assumed to be grid-following converters and not to provide any inertial response nor fast-frequency regulation.  All simulations are obtained with the software tool Dome \cite{dome}.

\subsection{Fault near a \gls{DC}}
\label{sub:fault}

The case studies presented focus on a \gls{DC} with \qty{300} MW of installed server capacity which is being cooled by a \qty{60} MW chiller. This \gls{DC} is located in the outskirts of city of Dublin in an industrial area surrounded by generation power plants. It has been set that the internal \gls{DC} \gls{UPS} will trip if a voltage variation of \qty{\pm0.1}{\perunit} and a frequency variation of \qty{\pm0.3}  Hz is observed. The disconnection time of the \gls{DC} is equal to \qty{30} s. The \gls{DC} load profile is considered as constant.

\begin{figure}[!htb]
\centering
\includegraphics[width=0.95\columnwidth]{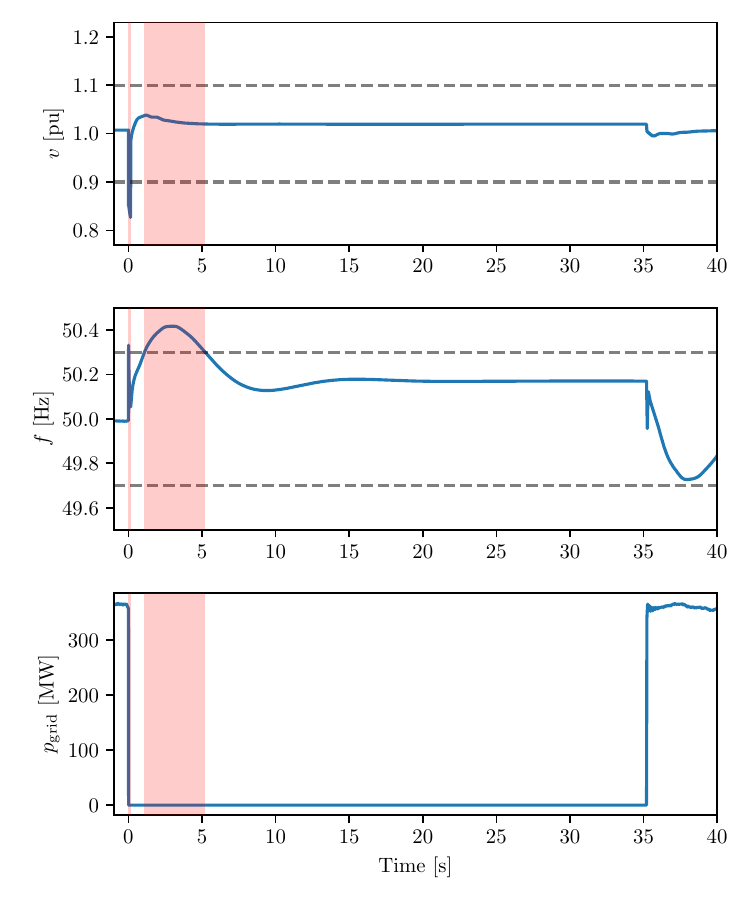}
\caption{\gls{DC} under a fault considering a constant load.}
\label{fig:Datacenter_Fault}
\vspace{-3mm}
\end{figure}

We simulate a fault in a bus near the \gls{DC}.  Fig.~\ref{fig:Datacenter_Fault} shows the magnitude and frequency of the voltage at its terminals, with the \gls{DC} load demand shown as well in the third subplot.  The shaded area in red corresponds to the time the external grid parameters of the grid as measured by the  \gls{UPS} are out of bounds. The \gls{UPS} detects an undervoltage and quickly disconnects the \gls{DC} from the grid with no delay. This sudden drop in load causes an imbalance in the generation versus the load that rises the frequency. After the transient is cleared, the \gls{UPS} starts counting \qty{30} s and finally reconnects the \gls{DC} to the rest of the grid.

\subsection{Flapping}

In this scenario, we consider a fast reconnection scheme, that is, we consider the case for which the \gls{DC} is reconnected 10 s after the conditions at the point-of-connection bus of the grid recover normal operating conditions.  

\begin{figure}[!htb]
\centering
\includegraphics[width=0.95\columnwidth]{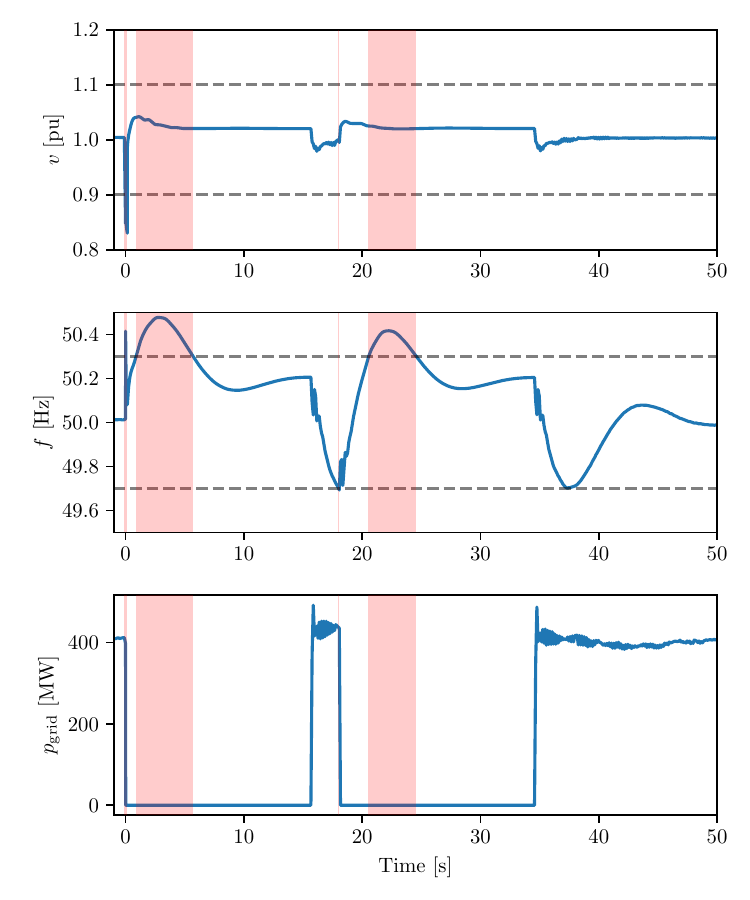}
\caption{\gls{DC} flapping (disconnection time = \qty{10}{\s}).}
\label{fig:Datacenter_Flapping_10}
\vspace{-3mm}
\end{figure}

Fig.~\ref{fig:Datacenter_Fault} shows that \glspl{DC} provoke large transients in the grid owing to their large power consumption and sudden disconnection from the grid.  This figure also shows that the frequency of grid goes down during the reconnection of the \gls{DC} close to the disconnection threshold of the \gls{DC}.  If the \gls{DC} is larger, it can generate flapping \cite{chatzivasileiadisMicroflexibilityChallengesPower2023}.  The reconnection triggers, leading to a frequency drop below \qty{49.7} Hz, disconnecting the \gls{DC} again.  Fig.~\ref{fig:Datacenter_Flapping_10} shows this phenomenon.  To obtain these results, we have assumed that the \gls{DC} has double cooling load with respect to the previous scenario.  The total power consumption of the \gls{DC} is thus around \qty{420} MW.  Setting the disconnection time equal to \qty{10} s shows that the sudden disconnection pushes the frequency at the \gls{DC} terminals below \qty{49.7} Hz, re-disconnecting the \gls{DC}.  The cycle can repeat indefinitely in the worst scenario.  In this case, the frequency during the next reconnection is just slightly above \qty{49.7} Hz, exiting the loop.  

Flapping can be avoided in various ways.  An obvious option is to increase the time the \gls{DC} waits before reconnecting to the grid.  In the scenario discussed in Section \ref{sub:fault} the reconnection time is 30 s which effectively prevents the occurrence of the flapping.  However, in general, it is not possible \textit{a priori} to know what reconnection time will cause the flapping.  Moreover, for same reconnection time, the risk of flapping is higher the higher the size of the \glspl{DC} and the lower the current loading condition of the grid.  

\begin{figure}[!htb]
\centering
\includegraphics[width=0.95\columnwidth]{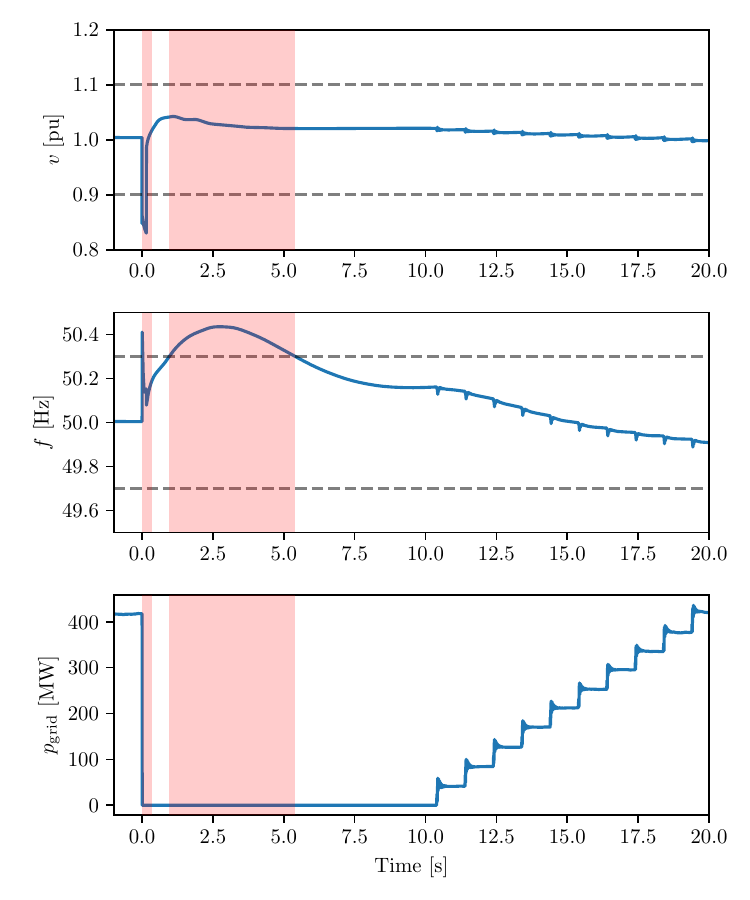}
\caption{\gls{DC} gradual reconnection.}
\label{fig:Datacenter_Gradual_Reconnection}
\vspace{-3mm}
\end{figure}

Another, more sustainable and safer option from the point of view of the grid is the segmentation of the \gls{DC} facility into different \glspl{UPS} with different reconnection times.  Fig.~\ref{fig:Datacenter_Gradual_Reconnection} has the \gls{DC} considered for Fig.~\ref{fig:Datacenter_Flapping_10} and splits the load into 10 different \gls{UPS}.  The quickest \gls{UPS} has a reconnection time of \qty{5} s, while the next one is \qty{1} s slower, and successively longer reconnection times for the remaining ones.  This segmentation avoids the flapping issue, enabling a seamless transition from fully disconnection to fully connection that avoids significant voltage and frequency transients in the grid.  This approach mirrors the ramp of reconnecting \glspl{DC} observed in the Irish system \cite{kerciDynamicResponseInverterbased2023}.

\vspace{-2mm}
\subsection{Batched and \gls{AI} Loads}

\begin{figure}[!htb]
	\centering
	\includegraphics[width=0.95\columnwidth]{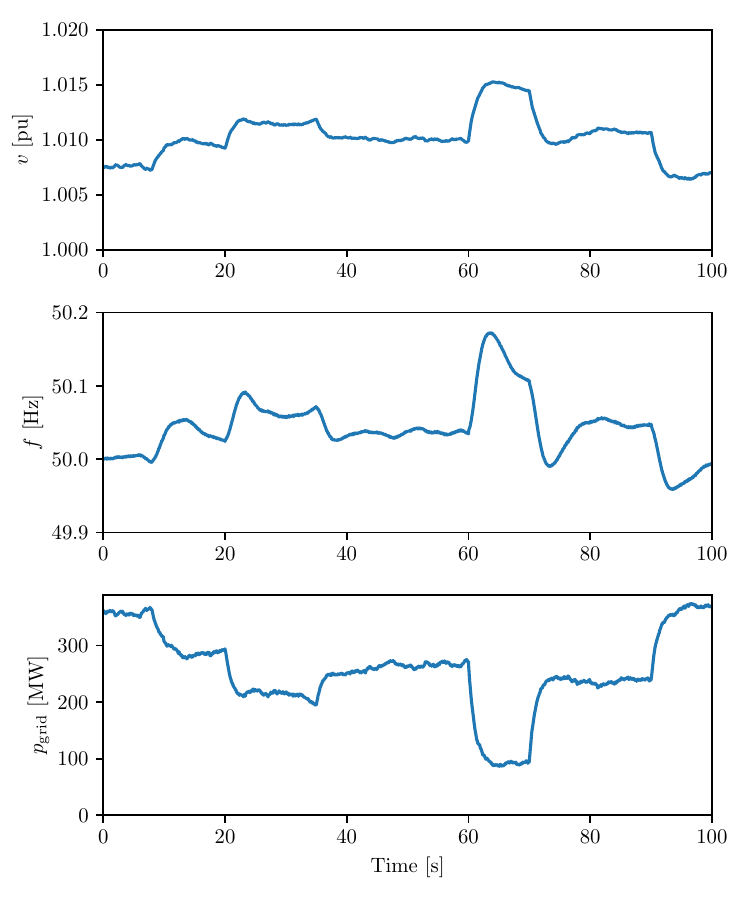}
	\caption{\gls{DC} CPU batched load.}
	\label{fig:Datacenter_Batched_Load}
\vspace{-3mm}
\end{figure}

\glspl{DC} can cause transients not only due to their behavior during and after a fault, but also during their normal functioning. As an example, a \gls{DC} with \qty{300}  MW of installed server capacity and \qty{60}  MW of cooling that works with a batched load pattern shows the voltage, frequency and power variations presented in Fig.~\ref{fig:Datacenter_Batched_Load}.
Each task that the \gls{DC} ought to compute is queued and executed accordingly. It is quite possible that some tasks take advantage of a lower number of individual servers, which explains the power consumption variations. The duration and load percentage of the \gls{DC} is arbitrary. For this graph, it was plotted a set of tasks whose duration was $\approx$\qty{25} s and the usage varied from \qty{10}{\%} to \qty{100}{\%} of the \gls{DC} total capacity.

\begin{figure}[!htb]
\centering
\includegraphics[width=0.95\columnwidth]{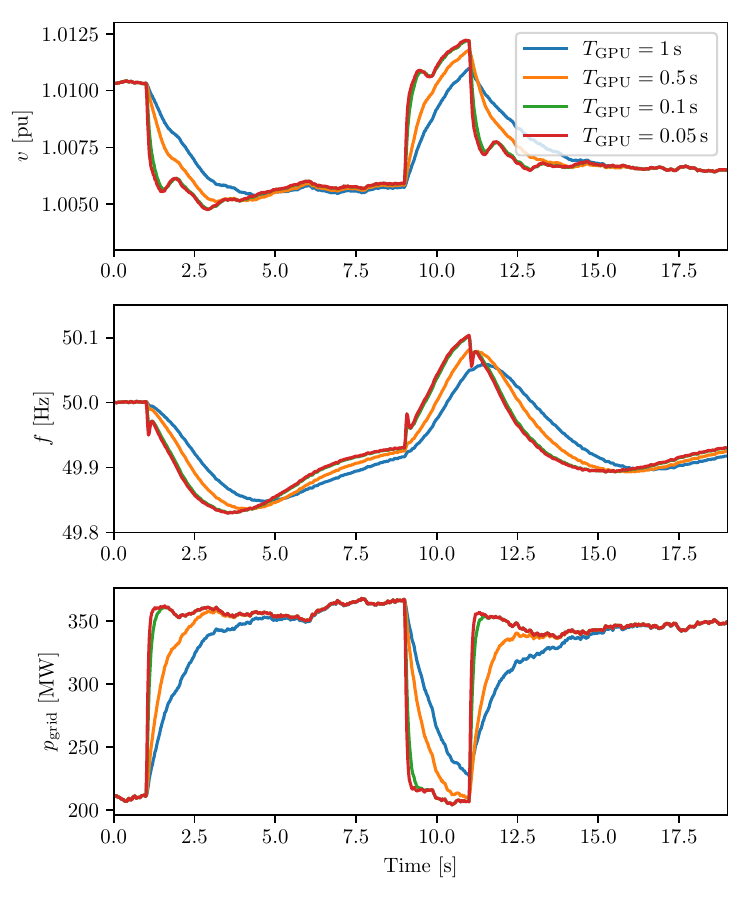}
\caption{\gls{DC} \gls{AI} load.}
\label{fig:Datacenter_Ai_Load}
\vspace{-4mm}
\end{figure}

These variations play a more significant role for the \gls{AI} \glspl{DC}.  Fig.~\ref{fig:Datacenter_Ai_Load} shows a scenario where the \qty{300} MW of the \gls{DC} are divided into \qty{150}  MW of constant \gls{CPU} load and \qty{150}  MW of \gls{GPU} load.  Each pulse of the GPU load has been set to \qty{10} s, with duty cycle of 0.8, i.e., \qty{80}{\%} of the time of the pulse the \glspl{GPU} are connected and drawing power, while \qty{20}{\%} of the time they are idle.  Results indicate a large frequency variation, while the voltage drops about \qty{0.004}{\perunit}.  These variations might cause issues in the grid, but they can be dampened by properly adjusting the $T_{\GPU}$ parameter. A larger time constant  dampens the oscillations caused by sudden power consumption increases, such as the ones caused by \gls{AI} \glspl{DC}.

\vspace{-2mm}
\subsection{Periodic Transients}

\begin{figure}[!htb]
\centering
\includegraphics[width=0.95\columnwidth]{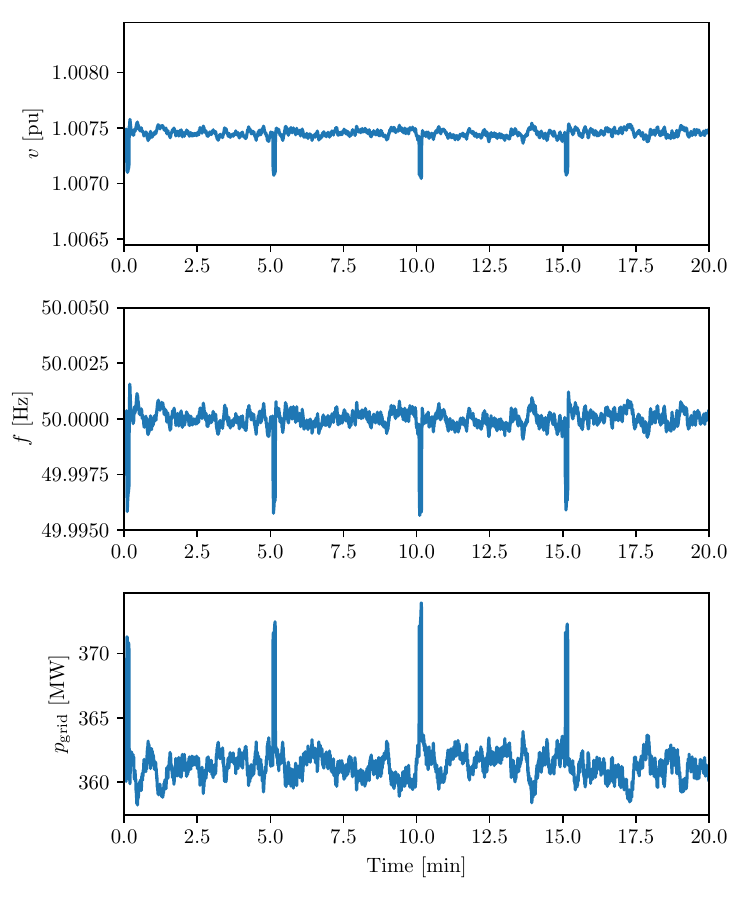}
\caption{\gls{DC} \gls{CPU} cyclic transients.}
\label{fig:Cyclic_Transients}
\end{figure}

\glspl{DC}, as noted in \cite{kerciImpactConverterbasedDemand2024}, have spikes of demand which are regular in nature.  Fig.~\ref{fig:Cyclic_Transients} shows the effect on the Irish system of an increase of 10 MW of demand every 5 minutes in a \gls{DC}.  These spikes create micro-transients both for voltage and frequency.  These transients are small compared with those caused by faults and sudden reconnections of the \gls{DC}, but nevertheless concerns the Irish TSO as they can reduce the frequency quality (e.g., minutes outside the $\pm$100 mHz band).  These periodic transients are due to regularly scheduled tasks that the servers within the \gls{DC} perform.  As \glspl{DC} consist of a myriad of individual devices, 
de-synchronizing these tasks can effectively remove the demand spikes.

\subsection{Detailed Reconnection}

\begin{figure}[!htb]
\centering
\includegraphics[width=0.95\columnwidth]{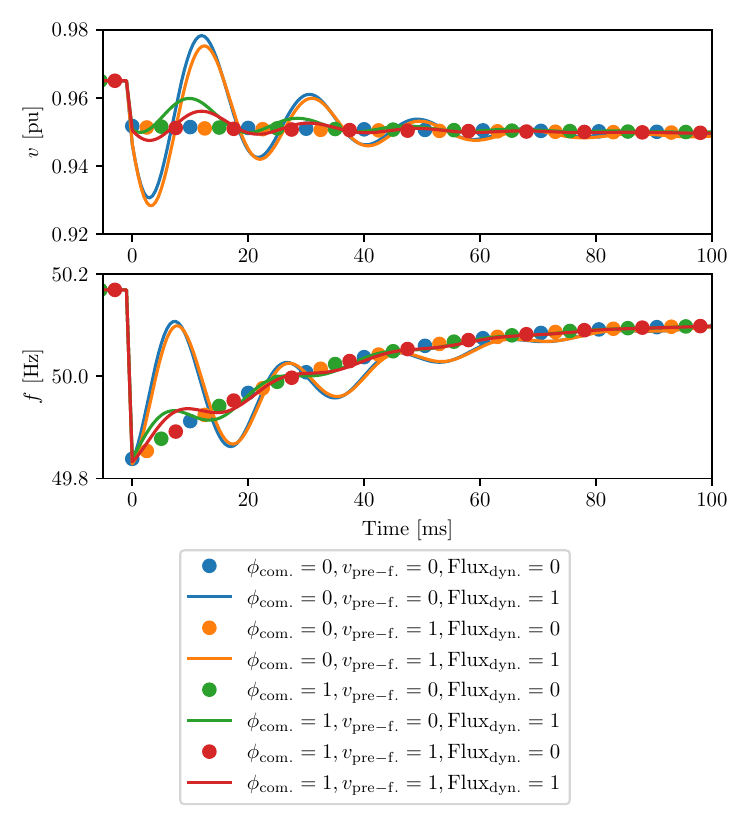}
\caption{Reconnection of a \gls{DC} under a fault with constant load.}
\label{fig:Datacenter_Fault_Detailed}
%\vspace{-3mm}
\end{figure}

In this last scenario, we consider a detailed reconnection model, including  the effect of the dynamics of the induction motor utilized for the cooling system.  Fig.~\ref{fig:Datacenter_Fault_Detailed} shows the values of voltage and frequency at the \gls{DC} terminals during the reconnection when considering (or not) angle compensation ($\phi_{\rm com.}$), using the pre-fault voltage ($v_{\rm pre-f.}$), and flux dynamics for the cooling load (${\rm Flux}_{\rm dyn.}$).  As in all previous cases the internal voltage used within the \gls{DC} while it was disconnected was read \qty{10} ms before the fault.  Angle compensation and flux dynamics of the induction motor that cools the \gls{DC} are also included in the model.

Results indicate that angle compensation can mitigate the fastest transients during the reconnection stage of the \gls{DC}.  On the other hand, setting the internal voltage using the pre-fault voltage or the fixed value \qty{1.0} pu does not much difference for this case study owing to the fact that the previous voltage ($\approx$\qty{0.95}{\perunit}) was already close to \qty{1.0} pu.  Fig.~\ref{fig:Datacenter_Fault_Detailed} also shows the results if flux dynamics are ignored (results shown with dots).  For this model angle compensation makes no difference as no internal loads use the voltage angle as an input.

\section{Conclusions} 
\label{sec:conclusions}

This paper proposes a \gls{DC} model that captures the time-varying power consumption and the response following faults and other grid disturbances.  The model is tested using a realistic model of the all-island transmission system and information on the \glspl{DC} that are currently installed in the system.  Main conclusions are summarized below.

\begin{itemize}
\item Traditional \glspl{DC} can be regarded as a constant load.  Nevertheless, the introduction of \gls{AI} \glspl{DC} have changed this dramatically.  Due to the way that they make use of \gls{AI} accelerators, mainly \glspl{GPU} and \glspl{TPU}, strong transients due to sudden losses of hundreds of MWs that occur.  These transients can be mitigated by adjusting the way the scaling drivers of \glspl{CPU} and other devices work.
\item Reconnection is a sensitive procedure that can destabilize, or push the grid into an unstable condition due to the sudden demand of large amounts of electric power.  The article shows that a \gls{DC} made out of \glspl{UPS} with different reconnection times allows for a seamless reconnection with only moderate transients.

\item Sudden changes in demand are caused by variations of the computing load of the \glspl{DC}.  The paper shows that these transients can be controlled and mitigated by \gls{DC} by adjusting the speed of the frequency changes.  This setting can be controlled in hardware and software and, as \gls{TSO} requirements tighten, \gls{DC} might require to tune these parameters for both compliance and maximum performance.

\item The cooling load, which is modeled as an induction generator, introduces transients during the reconnection stage if flux dynamics are considered.  Results show that matching the voltage grid angle by the \gls{UPS} dampen these transients, while maintaining the last voltage of the grid results in a much smaller effect.
\end{itemize}

Modeling of \gls{DC} for power system analysis is in an early stage.  Future work will focus on the modeling of \gls{DC} long-term behavior and explore control strategies that achieve gradual changes of the consumption of \glspl{DC}.

\section*{Acknowledgments}

The authors wish to thank Dr Taulant K\"er{\c{c}}i, EirGrid Group, for the providing information on data centers and their effect on the Irish system dynamic performance.

%\bibliographystyle{IEEEtran}
%\bibliography{references}{}

\begin{thebibliography}{10}
\providecommand{\url}[1]{#1}
\csname url@samestyle\endcsname
\providecommand{\newblock}{\relax}
\providecommand{\bibinfo}[2]{#2}
\providecommand{\BIBentrySTDinterwordspacing}{\spaceskip=0pt\relax}
\providecommand{\BIBentryALTinterwordstretchfactor}{4}
\providecommand{\BIBentryALTinterwordspacing}{\spaceskip=\fontdimen2\font plus
\BIBentryALTinterwordstretchfactor\fontdimen3\font minus
  \fontdimen4\font\relax}
\providecommand{\BIBforeignlanguage}[2]{{%
\expandafter\ifx\csname l@#1\endcsname\relax
\typeout{** WARNING: IEEEtran.bst: No hyphenation pattern has been}%
\typeout{** loaded for the language `#1'. Using the pattern for}%
\typeout{** the default language instead.}%
\else
\language=\csname l@#1\endcsname
\fi
#2}}
\providecommand{\BIBdecl}{\relax}
\BIBdecl

\bibitem{kerciImpactConverterbasedDemand2024}
T.~K{\"e}r{\c c}i, C.~Duggan, U.~Farooq, S.~Tweed, and M.~Val~Escudero,
  ``Impact of {{Converter-based Demand}} on {{Frequency Quality}} in the
  {{Ireland}} and {{Northern Ireland Power Systems}},'' in \emph{C4 {{PS1}} -
  {{Power}} System Dynamic Analysis in the Energy Transition: Challenges,
  Opportunities and Advances}, Paris, France, Aug. 2024, p.~11.

\bibitem{cru_2021}
CRU, ``{{CRU Direction}} to the {{System Operators}} related to {{Data Centre}}
  grid connection processing,'' Commission for Regulation of Utilities (Dublin,
  Ireland) - An Coimisi{\'u}n um Rial{\'a}il F{\'o}ntais (Baile {\'A}tha
  Cliath, {\'E}ire), Dublin, Ireland, Decision {{Paper}} CRU/21/124, Nov. 2021.

\bibitem{cupelliDataCenterControl2018}
L.~Cupelli, T.~Sch{\"u}tz, P.~Jahangiri, M.~Fuchs, A.~Monti, and D.~M{\"u}ller,
  ``Data {{Center Control Strategy}} for {{Participation}} in {{Demand Response
  Programs}},'' \emph{IEEE Trans.~on Industrial Informatics}, vol.~14, no.~11,
  pp. 5087--5099, Nov. 2018.

\bibitem{luoImpactDatacenterLoad2020}
P.~Luo, X.~Wang, and Y.~Li, ``The impact of datacenter load regulation on the
  stability of integrated power systems,'' \emph{Sustainable Energy
  Technologies and Assessments}, vol.~42, p. 100875, Dec. 2020.

\bibitem{fanPowerProvisioningWarehousesized2007}
X.~Fan, W.-D. Weber, and L.~A. Barroso, ``Power provisioning for a
  warehouse-sized computer,'' in \emph{ISCA}, New York, Jun. 2007, pp. 13--23.

\bibitem{zhangUnlockingFlexibilitiesData2025}
Y.~Zhang, H.~Tang, H.~Li, and S.~Wang, ``Unlocking the flexibilities of data
  centers for smart grid services: {{Optimal}} dispatch and design of energy
  storage systems under progressive loading,'' \emph{Energy}, vol. 316, p.
  134511, Feb. 2025.

\bibitem{habibikhalajEnergyEnvironmentalEconomical2016}
A.~Habibi~Khalaj, T.~Scherer, and S.~K.~Halgamuge, ``Energy, environmental and
  economical saving potential of data centers with various economizers across
  {{Australia}},'' \emph{Applied Energy}, vol. 183, pp. 1528--1549, Dec. 2016.

\bibitem{chuResearchStatusDevelopment2023}
J.~Chu and X.~Huang, ``Research status and development trends of evaporative
  cooling air-conditioning technology in data centers,'' \emph{Energy and Built
  Environment}, vol.~4, no.~1, pp. 86--110, Feb. 2023.

\bibitem{beghiModellingControlFree2017}
A.~Beghi, L.~Cecchinato, G.~D. Mana, M.~Lionello, M.~Rampazzo, and E.~Sisti,
  ``Modelling and control of a free cooling system for {{Data Centers}},''
  \emph{Energy Procedia}, vol. 140, pp. 447--457, Dec. 2017.

\bibitem{anDynamicCouplingRealtime2022}
H.~An and X.~Ma, ``Dynamic coupling real-time energy consumption modeling for
  data centers,'' \emph{Energy Reports}, vol.~8, pp. 1184--1192, Nov. 2022.

\bibitem{fuEquationbasedObjectorientedModeling2019}
Y.~Fu, W.~Zuo, M.~Wetter, J.~W. VanGilder, and P.~Yang, ``Equation-based
  object-oriented modeling and simulation of data center cooling systems,''
  \emph{Energy and Buildings}, vol. 198, pp. 503--519, Sep. 2019.

\bibitem{hamSimplifiedServerModel2015}
S.-W. Ham, M.-H. Kim, B.-N. Choi, and J.-W. Jeong, ``Simplified server model to
  simulate data center cooling energy consumption,'' \emph{Energy and
  Buildings}, vol.~86, pp. 328--339, Jan. 2015.

\bibitem{drovtarUtilizingDemandResponse2025}
I.~Drovtar, M.~Leinakse, K.~Tuttelberg, and J.~Kilter, ``Utilizing {{Demand
  Response}} in {{Load Modelling}} for {{Voltage}} and {{Reactive Power Control
  Studies}},'' \emph{IEEE Trans.~on Power Systems}, vol.~40, no.~2, pp.
  1389--1400, Mar. 2025.

\bibitem{suryanarayanaSystemModelingFault2021}
H.~Suryanarayana \emph{et~al.}, ``System modeling and fault studies in data
  center power distribution,'' in \emph{CIRED}, vol. 2021, Sep. 2021, pp.
  1435--1439.

\bibitem{sunDynamicModelConverterBased2022}
J.~Sun, S.~Wang, J.~Wang, and L.~M. Tolbert, ``Dynamic {{Model}} and
  {{Converter-Based Emulator}} of a {{Data Center Power Distribution
  System}},'' \emph{IEEE Trans.~on Power Electronics}, vol.~37, no.~7, pp.
  8420--8432, Jul. 2022.

\bibitem{dayarathnaDataCenterEnergy2016}
M.~Dayarathna, Y.~Wen, and R.~Fan, ``Data {{Center Energy Consumption
  Modeling}}: {{A Survey}},'' \emph{IEEE Communications Surveys \& Tutorials},
  vol.~18, no.~1, pp. 732--794, 2016.

\bibitem{cheungSimplifiedPowerConsumption2018}
H.~Cheung, S.~Wang, C.~Zhuang, and J.~Gu, ``A simplified power consumption
  model of information technology ({{IT}}) equipment in data centers for energy
  system real-time dynamic simulation,'' \emph{Applied Energy}, vol. 222, pp.
  329--342, Jul. 2018.

\bibitem{ahmedElectricalEnergyConsumption2019}
K.~M.~U. Ahmed, J.~Sutaria, M.~H.~J. Bollen, and S.~K. R{\"o}nnberg,
  ``Electrical {{Energy Consumption Model}} of {{Internal Components}} in
  {{Data Centers}},'' in \emph{{ISGT-Europe}}, Sep. 2019, pp. 1--5.

\bibitem{pourbeikAggregateDynamicModel2019}
P.~Pourbeik, J.~Weber, D.~Ramasubramanian, J.~{Sanchez-Gasca}, J.~Senthil,
  P.~Zadkhast, J.~Boemer, A.~Gaikwad, I.~Green, S.~Tacke, R.~Favela, S.~Wang,
  and S.~Zhu, ``An aggregate dynamic model for distributed energy resources for
  power system stability studies,'' 2019.

\bibitem{chandrakasanMinimizingPowerConsumption1995}
A.~Chandrakasan and R.~Brodersen, ``Minimizing power consumption in digital
  {{CMOS}} circuits,'' \emph{Proceedings of the IEEE}, vol.~83, no.~4, pp.
  498--523, Apr. 1995.

\bibitem{russellArtificialIntelligenceModern2016}
S.~Russell and P.~Norvig, \emph{Artificial {{Intelligence}}: {{A Modern
  Approach}}, {{Global Edition}}}, 3rd~ed.\hskip 1em plus 0.5em minus
  0.4em\relax Boston: Pearson, May 2016.

\bibitem{ryanAssessmentLargeLoad2025}
Q.~Ryan, J.~Zhao, and K.~Thomas, ``An {{Assessment}} of {{Large Load
  Interconnection Risks}} in the {{Western Interconnection}},'' Western
  Electricity Coordinating Council (WECC), Tech. Rep., Feb. 2025.

\bibitem{chenDataCenterPower2023}
Y.~Chen, K.~Shi, M.~Chen, and D.~Xu, ``Data {{Center Power Supply Systems}}:
  {{From Grid Edge}} to {{Point-of-Load}},'' \emph{IEEE J.~of Emerging and
  Selected Topics in Power Elec.}, vol.~11, no.~3, pp. 2441--2456, Jun. 2023.

\bibitem{barrosoDatacenterComputerDesigning2019}
L.~A. Barroso, U.~H{\"o}lzle, and P.~Ranganathan, \emph{The {{Datacenter}} as a
  {{Computer}}: {{Designing Warehouse-Scale Machines}}}.\hskip 1em plus 0.5em
  minus 0.4em\relax Springer, 2019.

\bibitem{gudrun:2019}
G.~M. J{\'o}nsd{\'o}ttir and F.~Milano, ``Modeling solar irradiance for
  short-term dynamic analysis of power systems,'' in \emph{IEEE PES General
  Meeting}, 2019, pp. 1--5.

\bibitem{rafael:2013}
F.~Milano and R.~Zárate-Miñano, ``A systematic method to model power systems
  as stochastic differential algebraic equations,'' \emph{IEEE Trans.~on Power
  Systems}, vol.~28, no.~4, pp. 4537--4544, 2013.

\bibitem{oliveiraComparativePerformanceAnalysis2019}
F.~Oliveira and A.~Ukil, ``Comparative {{Performance Analysis}} of
  {{Induction}} and {{Synchronous Reluctance Motors}} in {{Chiller Systems}}
  for {{Energy Efficient Buildings}},'' \emph{IEEE Trans.~on Industrial
  Informatics}, vol.~15, no.~8, pp. 4384--4393, Aug. 2019.

\bibitem{zhangSurveyDataCenter2021}
Q.~Zhang \emph{et~al.}, ``A survey on data center cooling systems:
  {{Technology}}, power consumption modeling and control strategy
  optimization,'' \emph{J.~of Systems Architecture}, vol. 119, p. 102253, Oct.
  2021.

\bibitem{hasanuzzamanEnergySavingsEmissions2011}
M.~Hasanuzzaman, N.~A. Rahim, R.~Saidur, and S.~N. Kazi, ``Energy savings and
  emissions reductions for rewinding and replacement of industrial motor,''
  \emph{Energy}, vol.~36, no.~1, pp. 233--240, Jan. 2011.

\bibitem{kundurPowerSystemStability1994}
P.~Kundur, \emph{{Power System Stability and Control}}.\hskip 1em plus 0.5em
  minus 0.4em\relax New York: MHP, 1994.

\bibitem{nercincident2025}
{North American Electric Reliability Corporation (NERC)}, ``Incident {{Review}}
  - {{Considering Simultaneous Voltage-Sensitive Load Reductions}},'' NERC,
  USA, Incidence, Jan. 2025.

\bibitem{DataCenterMap}
``Data {{Center Map}} - {{Colocation}}, {{Cloud}} and {{Connectivity}},''
  https://www.datacentermap.com/.

\bibitem{linAdaptingDatacenterCapacity2023}
L.~Lin and A.~A. Chien, ``Adapting {{Datacenter Capacity}} for {{Greener
  Datacenters}} and {{Grid}},'' in \emph{Proceedings of the 14th {{ACM
  International Conference}} on {{Future Energy Systems}}}, ser. E-{{Energy}}
  '23.\hskip 1em plus 0.5em minus 0.4em\relax New York, NY, USA: Association
  for Computing Machinery, Jun. 2023, pp. 200--213.

\bibitem{dome}
F.~Milano, ``A {Python}-based software tool for power system analysis,'' in
  \emph{IEEE PES General Meeting}, 2013, pp. 1--5.

\bibitem{chatzivasileiadisMicroflexibilityChallengesPower2023}
S.~Chatzivasileiadis \emph{et~al.}, ``Micro-flexibility: {{Challenges}} for
  power system modeling and control,'' \emph{Electric Power Systems Research},
  vol. 216, p. 109002, Mar. 2023.

\bibitem{kerciDynamicResponseInverterbased2023}
T.~K{\"e}r{\c c}i, ``Dynamic {{Response}} of {{Inverter-based Resources}} in
  the {{Ireland}} and {{Northern Ireland Power Systems}},'' in \emph{2013 IEEE
  Belgrade PowerTech}, Belgrade, Jun. 2023, pp. 1--6.

\end{thebibliography}

% Generated by IEEEtran.bst, version: 1.14 (2015/08/26)

\vfill
\end{document}